%% file: paper.tex
\documentclass[conference]{IEEEtran}
\IEEEoverridecommandlockouts

\usepackage[top=0.75in,bottom=1.0in,left=0.625in,right=0.625in]{geometry}
\usepackage[utf8]{inputenc} 
\usepackage[T1]{fontenc}
\usepackage{url}
\usepackage{ifthen}
\usepackage{cite}
\usepackage{verbatim}
\usepackage[cmex10]{amsmath} 
\usepackage{makecell}
\usepackage{textcase}
\usepackage[tablename=Table]{caption}
\usepackage{blindtext}
\usepackage{floatrow}
\usepackage{soul}
\usepackage{gensymb}
\usepackage{cite}
\usepackage{amsmath,amssymb,amsfonts}
\usepackage{textcomp}
\usepackage{graphicx}
\usepackage{epsfig}
\usepackage{color}
\usepackage{fancybox} 
\usepackage{multirow}
\usepackage{setspa ce}
\usepackage{psfrag}
\usepackage{booktabs}
\usepackage{float}
\usepackage{algorithm}
\usepackage{algpseudocode}
\usepackage{mathtools, nccmath, bigints}
\usepackage{subfigure} 
\usepackage{caption}
\usepackage{lipsum}
\usepackage{mathrsfs}								

\usepackage{placeins}
\newcommand\blfootnote[1]{%
  \begingroup
  \renewcommand\thefootnote{}\footnote{#1}%
  \addtocounter{footnote}{-1}%
  \endgroup
}

\newfloatcommand{capbtabbox}{table}[][0.4\textwidth]

\input macro

\def\BibTeX{{\rm B\kern-.05em{\sc i\kern-.025em b}\kern-.08em
	T\kern-.1667em\lower.7ex\hbox{E}\kern-.125emX}}
\begin{document}

\title{Graph-based Untrained Neural Network Detector for OTFS Systems}

\author{%
	\IEEEauthorblockN{
		Hao Chang, 
        Branka Vucetic,
        and Wibowo Hardjawana,  
		\\
	}
		Centre for IoT and Telecommunications, The University of Sydney, Sydney, Australia.  \\
		\{hao.chang,branka.vucetic,wibowo.hardjawana\}@sydney.edu.au}

\maketitle
 
\begin{abstract} 
Inter-carrier interference (ICI) caused by mobile reflectors significantly degrades the conventional orthogonal frequency division multiplexing (OFDM) performance in high-mobility environments. The orthogonal time frequency space (OTFS) modulation system effectively represents ICI in the delay-Doppler domain, thus significantly outperforming OFDM. Existing iterative and neural network (NN) based OTFS detectors suffer from high complex matrix operations and performance degradation in untrained environments, where the real wireless channel does not match the one used in the training, which often happens in real wireless networks. In this paper, we propose to embed the prior knowledge of interference extracted from the estimated channel state information (CSI) as a directed graph into a decoder untrained neural network (DUNN), namely graph-based DUNN (GDUNN). We then combine it with Bayesian parallel interference cancellation (BPIC) for OTFS symbol detection, resulting in GDUNN-BPIC. Simulation results show that the proposed GDUNN-BPIC outperforms state-of-the-art OTFS detectors under imperfect CSI.   
\end{abstract}

\blfootnote{Part of this work has been published as an MPhil thesis in \cite{hcthesis}.}
\section{Introduction}  
In high-mobility environments, Doppler shifts lead to high inter-carrier interference (ICI) in the 5G air interface that is based on orthogonal frequency division multiplexing (OFDM) system \cite{hadani2017orthogonal} and \cite{raviteja2018interference}. 
The orthogonal time frequency space (OTFS) proposed in \cite{hadani2017orthogonal} as a 6G air interface addresses ICI by multiplexing symbols in the delay-Doppler (DD) domain. The DD domain captures the geometry of the channel paths generated by mobile reflectors (e.g., delay and Doppler shift) and thus tracks the ICI, allowing the construction of a time-invariant channel matrix \cite{raviteja2018interference} for detectors. 
Multiple OTFS detectors \cite{9492800,9569353,long2021low,10154051,9900413} in the literature can be categorized as 1) iterative and 2) neural network (NN) based detectors.

Iterative OTFS detectors with good symbol error rate (SER) performances have been proposed in \cite{9492800} and \cite{9569353}, referred to as unitary approximate message passing (UAMP) and Bayesian parallel interference cancellation with minimum-mean-square-error (MMSE-BPIC) detectors, respectively.  
UAMP performs unitary transformation operations on the OTFS channel matrix and received signal using a singular value decomposition (SVD). The results of this transformation are used in the AMP algorithm \cite{9492800} to estimate transmitted OTFS symbols iteratively. 
MMSE-BPIC first performs a matrix inversion operation to compute MMSE symbol estimates, which are used as initial symbol estimates of BPIC in the first iteration. Throughout the iterative process, the PIC scheme is used to subtract interference from the received signal using the symbol estimates from the previous iteration. The Bayesian concept is applied to estimate the mean and variance of posterior symbols, and these estimates are weighted with their estimation in the previous iteration through decision statistic combining (DSC).  
A near-optimal SER performance that is close to the one of maximum likelihood (ML) detector is achieved using the expectation propagation (EP) \cite{long2021low} scheme. EP uses a matrix inversion and Bayesian concept to estimate the mean and variance of posterior symbols iteratively.
As the granularity of delay and Doppler shift to improve the tracking accuracy of the channel paths increases, the size of the DD domain time-invariant channel matrix also increases. This poses a challenge in complex matrix operations (e.g., SVD and matrix inversion) for the above detectors.  

NN based detectors can be categorized as 1) trained
NN and 2) untrained NN (UNN). Trained NN detectors \cite{10154051} and \cite{9900413} require a training process with a large amount of training data before implementation. Also, these detectors rely on the fidelity of the training data. The performance degradation will happen in untrained channel environments where the real wireless channel does not match the one used in the training, which often happens in real wireless networks.
UNN was first proposed in \cite{ulyanov2018deep} for image inverse problems without training, which uses an auto-encoder (AE) NN architecture consisting of an NN encoder followed by an NN decoder. The encoder and decoder consist of NN layers with decreasing and increasing sizes, respectively.
The AE used in \cite{ulyanov2018deep} consists of tens of layers for solving the image inverse problem iteratively, resulting in high computational complexity.
Our recently proposed decoder UNN OTFS detector, in combination with the BPIC scheme \cite{10225901}, referred to as DUNN-BPIC, uses only the decoder part of the AE to reduce the computational complexity. It performs symbol detection without any training and achieves a close to EP SER performance in the presence of \textit{perfect channel state information (CSI)} at the receiver. However, this is at the cost of large iteration numbers. The authors in \cite{9959969} embed prior knowledge as an undirected graph into a decoder UNN design and use it for iterative signal denoising. Downsampling of the graph representing prior knowledge to create a sub-graph is required to cater for decoder UNN layers with different sizes. This graph processing adds computational complexity. 
To date, the graph-based decoder UNN concept has not been applied in the OTFS systems under imperfect CSI estimates.
  
In this paper, we propose a graph-based decoder UNN (GDUNN) detector in the presence of imperfect CSI estimates at the OTFS receiver, in contrast to the perfect CSI assumption in \cite{10225901}.
We first transmit both pilot and data symbols at the transmitter. The symbols are then passed through a multi-path channel and impacted by the Doppler shift and channel delay.
The imperfect CSI estimates due to the receiver noise are obtained by using the transmitted pilot at the receiver.
The estimated CSI is then used to construct a directed graph representing the interference among the data symbols for GDUNN.
The directed graph is used due to a non-symmetrical channel matrix caused by different delays and Doppler shifts of different channel paths, in contrast to an undirected graph UNN in \cite{9959969} for a symmetric matrix. 
We add an additional NN layer after the DUNN output layer. The input of this layer is designed to match the matrix representation of the directed graph, allowing the prior knowledge of interference to be embedded without downsampling in contrast to \cite{9959969}. 
We then use the GDUNN output as the initial symbol estimates of BPIC for symbol detection, resulting in GDUNN-BPIC. 
The simulation results show that the proposed GDUNN needs significantly fewer iterations than DUNN. Furthermore, GDUNN-BPIC outperforms EP and other state-of-the-art OTFS detectors in the presence of \textit{imperfect CSI} using lower complexity.

{\bf Notations}: $a$, $\qa$, and $\qA$ denote scalar, vector, and matrix, respectively. $\qA^T$ represents the transpose of matrix $\qA$. 
$\mathbb{C}^{M\times N}$ and $\mathbb{R}^{M\times N}$ denote an $M\times N$ dimensional complex-valued and real-valued matrix, respectively. $\qI_M$ represents an $M\times M$ dimensional identity matrix. 
$[\cdot]_M$ represents the mod-$M$ operation.
The Kronecker product is denoted as $\otimes$. 
$\|\qx\|$ denotes the Frobenius norm of vector $\qx$. 
We use $\mathcal{N}(x:\mu,\sigma^2)$ to express a single variate Gaussian distribution of a random variable $x$ with mean $\mu$ and variance $\sigma^2$. 
We define $\qa = {\sf vec}(\qA)$ as the column-wise vectorization of matrix $\qA$, and $\qA = {\sf vec}^{-1}(\qa)$ denotes the vector elements folded back into a matrix. ${\sf diag}(\cdot)$ denotes a diagonalization operation on a vector.

\begin{figure}
	\centering
	\includegraphics[width=0.95\textwidth]{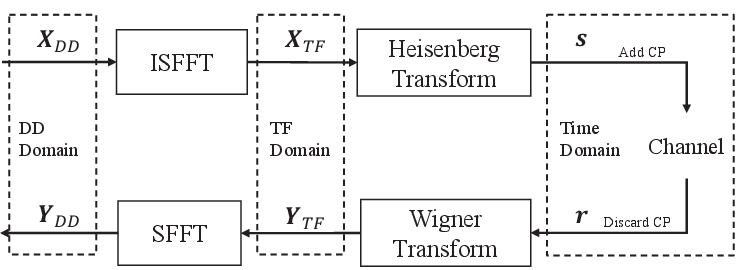}
	\caption{OTFS modulation system}
	\label{fig:sys-mod-general}
\end{figure} 

\begin{figure} 
\centering
\subfigure[$\qX_{\rm {DD}}$]
{\includegraphics[width=0.49\textwidth]{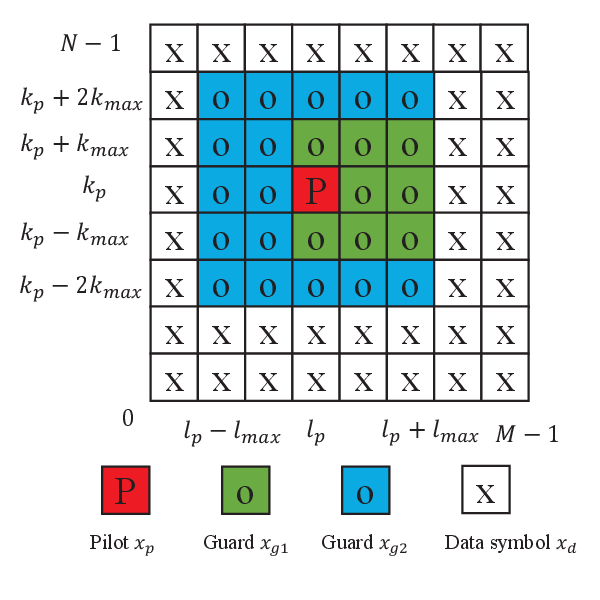}\hfill
\label{fig_xdd}}
\subfigure[$\qY_{\rm {DD}}$]
{\includegraphics[width=0.49\textwidth]{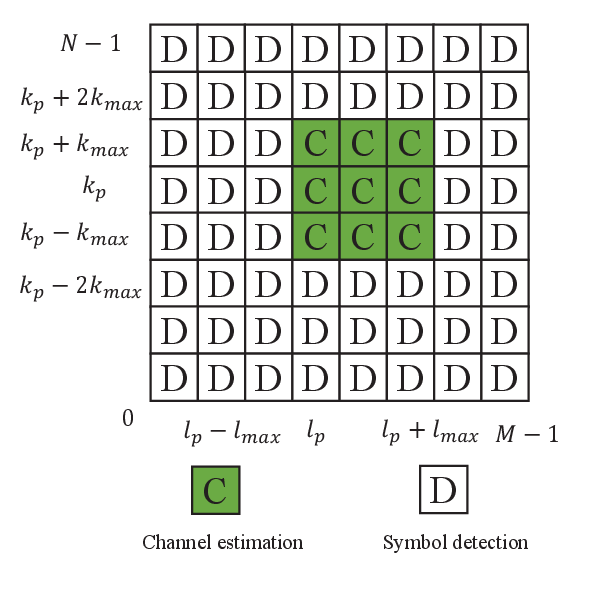}\hfill
\label{fig_ydd}}
\caption{An example of the DD domain OTFS frames for $k_p=l_p=3, N=M=8, k_{max}=1,l_{max}=2$}
\end{figure}\label{fig_dd}

\section{OTFS system model}\label{sys_mod}
In this section, we consider the OTFS system model and formulation in \cite{8516353,8671740}, as illustrated in Fig. \ref{fig:sys-mod-general}.
At the transmitter side, we first define a DD domain OTFS frame with $N$ Doppler shift and  $M$ delay indices, indicating the tracking of Doppler shift and delay at $\frac{k}{NT}$ and $\frac{l}{M\Delta f}$, $k=0,\dots,N-1, l=0,\dots,M-1$, respectively. $\Delta f$ is subcarrier spacing, and $T=\frac{1}{\Delta f}$ is the time slot duration.

Following \cite{8671740}, we then allocate the pilot, guard, and data symbols in the DD domain, denoted as $\qX_{\rm DD} \in \mathbb{C}^{N\times M}$. $x[k,l]$ represents the $(k,l)$-th entry of $\qX_{\rm DD}$, namely the symbol transmitted in Doppler index $k$ and delay index $l$,
\begin{equation}\label{x_p}
    x[k,l] =   
    \begin{cases}  
    x_p &  k=k_p,l=l_p,\\
    x_{g1} &  \begin{aligned} k_p-k_{max} \leq k \leq  k_p+k_{max}, \\l_p \leq l \leq  l_p+l_{max},\end{aligned}\\
    x_{g2} &  \begin{aligned} k_p-2k_{max} \leq k \leq  k_p+2k_{max}, \\l_p-l_{max} \leq l <  l_p, or 
    \\  k_p-2k_{max} \leq k <  k_p-k_{max},or, \\ k_p+k_{max} < k \leq  k_p+2k_{max},  
    \\l_p \leq l \leq  l_p+l_{max},\end{aligned}\\
    x_d & \rm{otherwise},
    \end{cases}  
\end{equation}
where $x_p$ is the pilot following binary phase shift keying (BPSK) with power $P_p$, and $k_p$ and $l_p$ represent the pilot position in the DD domain frame. We only allocate one pilot in $\qX_{\rm DD}$, as using multiple pilots will result in interference among those pilots. 
The guard $x_{g1}=0$ is used to ensure the Doppler shift and delay of $\frac{\pm k_{max}\Delta f}{N}$ Hz and $\frac{l_{max} T}{M}$ second can be tracked by using pilot $x_p$. 
The guard $x_{g2}=0$ is used to ensure the transmission of data symbols $x_d$ does not interfere with the pilot symbol $x_p$ in the presence of the previously stated Doppler shift and delay due to the mobility of reflectors.  
$x_d$ is the ${\Tilde{M}}$-ary quadrature amplitude modulation (${\Tilde{M}}$-QAM) data symbols with unit average power, and ${\Tilde{M}}$ is the modulation order. $k_{max}$ and $l_{max}$ are the maximum Doppler and delay indices for the OTFS multi-path channel.
Here, we give an example of the DD domain OTFS frames in Fig. \ref{fig_xdd}.

$\qX_{\rm DD}$ is then transformed into the time-frequency (TF) domain by using the inverse symplectic finite Fourier transform (ISFFT)\cite{8516353}, expressed as
\begin{equation}
	\qX_{\rm TF} = \qF_M\qX_{\rm DD}\qF_N^{\qH},
	\label{isfft}
\end{equation}
where $\qF_M \in \mathbb{C}^{M\times M}$ and $\qF_N^{\qH} \in \mathbb{C}^{N\times N}$ are $M$-points DFT and $N$-points IDFT matrices, and the $(p,q)$-th entries of them are $(\frac{1}{\sqrt{M}}e^{-j2\pi pq/M})_{p,q=0,\cdots,M-1}$ and $(\frac{1}{\sqrt{N}}e^{j2\pi pq/N})_{p,q=0,\cdots,N-1}$, respectively. $\qX_{\rm TF} \in \mathbb{C}^{M\times N}$ are the transmitted symbols in a TF domain OTFS frame. 
The Heisenberg transform \cite{8516353} is then applied to $\qX_{\rm TF}$ to generate the time domain transmitted signal, given as
\begin{equation}\label{eq:sysddmod-tx-heisenberg}
    \qs = {\sf vec}(\qG_{\rm tx}\qF^{\qH}_M\qX_{\rm TF}) = (\qF^{\qH}_N\otimes \qG_{\rm tx})\qx_{\rm DD},    
\end{equation}
where $\qG_{\rm tx}={\sf diag}\left[g_{\rm tx}(0),g_{\rm tx}(\frac{T}{M}),\dots,g_{\rm tx}(\frac{(M-1)T}{M})\right]$. $g_{\rm tx}(t)$ is a rectangular waveform with a duration of $T$, leading to $\qG_{\rm tx}=\qI_M$ \cite{8516353}. 
$\qx_{\rm DD}=\sf{vec}(\qX_{\rm DD})$. 
We then insert the cyclic prefix (CP) of length $l_{max}$ at the beginning of $\qs$ to prevent interference between different OTFS frames. They are then sampled at $\frac{1}{M\Delta f}$ second and sent via a wireless channel. Thus, $\qs$ occupies a total bandwidth of $M\Delta f$ Hz and a duration of $NT$ seconds.

The impulse response of the time-varying multi-path channel in OTFS is modeled as \cite{8516353}
\begin{equation}\label{htv}
    h(\tau, v) = \sum_{i=1}^P h_i \delta(\tau - \tau_i)\delta(v - v_i),
\end{equation}
where $\delta(\cdot)$ is the Dirac delta function, $h_i \sim \mathcal{N}(0, 1/P)$ denotes the $i$-th path gain, and $P$ is the total number of propagation paths. The delay and Doppler shift for path $i$ are $\tau_i = \frac{l_iT}{M}$ and $v_i = \frac{k_i\Delta f}{N}$, respectively. 
Each mobile reflector contributes to one unique path with its own delay index $l_i \in [0,  l_{max}]$ and Doppler shift index $k_i\in [-k_{max}, k_{max}]$ for $i=1,\dots, P$. The maximum channel delay and Doppler shift indices are set to $l_{max}\leq M-1$ and $k_{max}\leq \frac{N}{2}$, respectively. This implies a maximum supported channel delay of $\frac{(M-1)T}{M}$ seconds and a maximum Doppler shift of $\frac{\Delta f}{2}$ Hz.  

At the receiver, the time domain received signal is first sampled at intervals of $\frac{1}{M\Delta f}$. 
After discarding the CP, the time domain received signal $r(n)$ is then expressed as \cite{8516353} 
\begin{equation}
	r(n) = \sum^P_i h_ie^{j2\pi\frac{(n-l_i)k_i}{NM}} s([n-l_i]_{NM})+w(n),
	\label{eq:sysmod-r(n)}
\end{equation}
where $n=0,\dots,MN-1$, and $w(n)$ is the noise with each entry following $\mathcal{N}(0, \sigma^2)$. 
We then can write the vector form of \eqref{eq:sysmod-r(n)} as
\begin{equation}
	\qr = \tilde{\qH}\qs + \qw,
	\label{eq:sysmod-r}
\end{equation}
where $\tilde{\qH} = \sum_{i=1}^P h_i \qI_{MN}(l_i) \qDelta^{k_i}$ is the time domain channel matrix. 
The use of CP results in a cyclic shift matrix $\qI_{MN}(l_i)$ \cite{8516353}, which is obtained by circularly left shifting the columns of $\qI_{MN}$ by $l_i$.
$\qDelta^{k_i} = {\sf diag}\left[z^{0}, z^{1}, \cdots, z^{MN - 1}\right]$ is the $MN\times MN$ Doppler shift diagonal matrix, where $z=e^{\frac{j2\pi k_i}{MN}}$. 
$\qw$ is the noise. 
The TF domain received signal $\qY_{\rm TF} \in \mathbb{C}^{N\times M}$ is obtained by applying the Wigner transform to $\qr$, shown as \cite{8516353}
\begin{equation}
	\qY_{\rm TF} = \qF_M\qG_{\rm rx}\qR,
	\label{eq:sysddmod-rx-wigner}
\end{equation}
where $\qG_{\rm rx}={\sf diag}\left[g_{\rm rx}(0),g_{\rm rx}(\frac{T}{M}),\dots,g_{\rm rx}(\frac{(M-1)T}{M})\right]$. $g_{\rm rx}(t)=g_{\rm tx}(t)$ results in $\qG_{\rm rx}=\qG_{\rm tx}=\qI_M$ \cite{8516353}. $\qR={\sf vec}^{-1}(\qr)$.
The DD domain received signal $\qY_{\rm DD} \in \mathbb{C}^{M\times N}$ is obtained by applying the SFFT to $\qY_{\rm TF}$ \cite{8516353}
\begin{align}\label{eq:sysmod-rx-sfft}
	\qY_{\rm DD}  =\qF^{\qH}_M \qY_{\rm TF} \qF_N,
\end{align}
and the $(k,l)$-th entry of $\qY_{\rm DD}$ is $y[k,l]$ for $k=0,\cdots,N-1, l=0,\cdots,M-1$.
We then substitute \eqref{eq:sysddmod-tx-heisenberg}, \eqref{eq:sysmod-r(n)}, and \eqref{eq:sysddmod-rx-wigner} in \eqref{eq:sysmod-rx-sfft} and obtain
\begin{equation}\label{otfs_old}
    \qy_{\rm DD} =\underbrace{(\qF_N \otimes \qG_{\rm rx})\tilde{\qH}(\qF^{\qH}_N \otimes \qG_{\rm tx})}_{\qH_{\rm DD}}  \qx_{\rm DD} + \qw_{\rm DD} ,
\end{equation}
where $\qy_{\rm DD}=\sf{vec}(\qY_{\rm DD})$, and $\qw_{\rm DD} =(\qF_N \otimes \qG_{\rm rx})\qw$ denotes the noise in the DD domain. $\qH_{\rm DD}$ is the real DD domain OTFS channel matrix.

\section{OTFS Channel estimation}\label{ce}
We use the channel estimation scheme in \cite{8671740}.  
Based on the guard $x_{g1}$ in \eqref{x_p}, only a part of the received symbols $y[k,l]$, for $k_p-k_{max} \leq k \leq k_p+k_{max}, l_p \leq l \leq l_p+l_{max}$, i.e., the $(k,l)$-th entry of $\qY_{\rm {DD}}$ in \eqref{eq:sysmod-rx-sfft} are used to estimate the path gain and the corresponding delay and Doppler indices.
The Doppler shift and channel delay experienced by the transmitted pilot $x_p$ depends on the wireless channel \eqref{htv}. Here, we give an example of $\qY_{\rm {DD}}$ in Fig. \ref{fig_ydd} based on the symbol transmission in Fig. \ref{fig_xdd}.

As shown in \eqref{otfs_old}, the noise will impact the channel estimation process explained below.
For each of $y[k,l] \geq \epsilon$, we detect there is a path $j, j=1,\dots,\hat{P}$ with Doppler index $\hat{k}_j=k-k_p$, delay index $\hat{l}_j=l-l_p$ and path gain $\hat{h}_j=\frac{y[k,l]}{x_p}e^{-j2\pi (l-\hat{l}_j)\hat{k}_j}$ \cite{8671740}.  
We set $\epsilon = 3\sigma $, where $\sigma^2= 10^{\frac{-SNR}{10}}$.
We can then obtain the estimated time domain channel matrix $\hat{\qH}$ and the estimated DD domain channel matrix $\hat{\qH}_{\rm {DD}}$ by using \eqref{eq:sysmod-r} and \eqref{otfs_old}, and setting $l_i=\hat{l}_j,k_i=\hat{k}_j, h_i=\hat{h}_j$ and $P=\hat{P}$.
$\hat{\qH}_{\rm {DD}} \neq \qH_{\rm {DD}}$ as there is always a channel estimation error due to the incorrect estimation of $\hat{k}_j, \hat{l}_j$, $\hat{h}_j$ and $\hat{P}$.  
 
\begin{figure}
\centering
\subfigure[GDUNN]
{\includegraphics[width=0.62\textwidth]{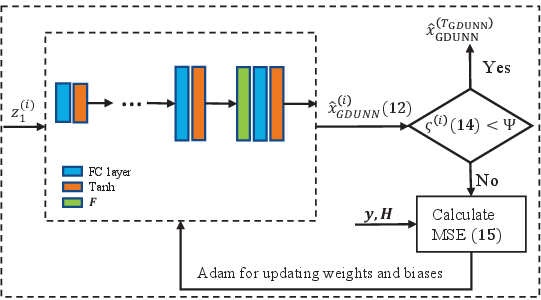}\hfill
\label{fig_NN}}
\subfigure[Directed graph]
{\includegraphics[width=0.35\textwidth]{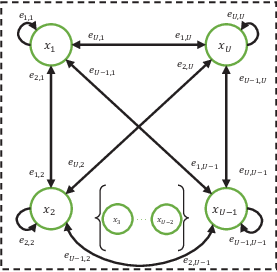}\hfill
\label{fig_graph}}
\caption{GDUNN architecture and graph design}
\end{figure}

\section{Graph-based untrained neural network symbol detector}
The OTFS transmission structure in \eqref{x_p} assumes the data symbols will be shifted by maximum $\pm k_{max}$ Doppler and $l_{max}$ delay indices. Thus, we will use all $y[k,l]$ in $\qY_{\rm {DD}}$ where $k\notin[k_p-k_{max},k_p+k_{max}], l\notin[l_p,l_p+l_{max}]$, which corresponds to the indices of $x_d$ and $x_{g2}$ in \eqref{x_p}.
This is equal to taking the $m$-th entry of $\qy_{\rm DD}$ in \eqref{otfs_old} where $m \neq (Np+q+1), p=l_p,\dots, l_p+l_{max},q= k_p-k_{max},\dots,k_p+k_{max}$. We write the received signal for data symbols as $\hat{\qy}_{\rm {DD}}$. 
The corresponding channel matrix for the data symbols can be written as $\hat{\qH}_{\rm {eff}}$, constructed by using the Doppler and delay indices of $x_d$ and $x_{g2}$ in \eqref{x_p}. This is equivalent to using the $m$-th row of $\hat{\qH}_{\rm {DD}}$, and $n$-th column of $\hat{\qH}_{\rm {DD}}$, where $n \neq (N\hat{p}+\hat{q}+1), \hat{p}=l_p-l_{max},\dots, l_p+l_{max},\hat{q}= k_p-2k_{max},\dots,k_p+2k_{max}$. Therefore, based on $\hat{\qy}_{\rm {DD}}$ and $\hat{\qH}_{\rm {eff}}$, model \eqref{otfs_old} for data detection can be rewritten as  
\begin{equation}\label{otfs_new}
    \hat{\qy}_{\rm {DD}} = \hat{\qH}_{\rm {eff}} \hat{\qx}_{\rm {DD}}+\hat{\qw}_{\rm {DD}}.
\end{equation}
where $\hat{\qx}_{\rm {DD}}$ consists of only the data symbols in $\qx_{\rm {DD}}$, i.e., the $n$-th entry of $\qx_{\rm {DD}}$. $\hat{\qw}_{\rm {DD}}$ is the noise, consisting of the relevant $m$-th entry of $\qw_{\rm {DD}}$.
$\hat{\qy}_{\rm {DD}}, \hat{\qw}_{\rm {DD}} \in \mathbb{C}^{\frac{V}{2}\times 1} $, $\hat{\qH}_{\rm {eff}}\in \mathbb{C}^{\frac{V}{2}\times \frac{U}{2}}$, and $\hat{\qx}_{\rm {DD}} \in \mathbb{C}^{\frac{U}{2}\times 1}$ where $V=2(NM-(l_{max}+1)(2k_{max}+1)), U=2(NM-(2l_{max}+1)(4k_{max}+1))$.   

We then consider a real-valued equivalent model of \eqref{otfs_new} as
\begin{align}
	\qy = \qH \qx + \qn,
	\label{real-y=hx+w}
\end{align}	 
where $\qx= \begin{bmatrix}
	\Re(\hat{\qx}_{\rm DD})  \ \Im(\hat{\qx}_{\rm DD})
\end{bmatrix}^T$,
$\qy=
\begin{bmatrix}
	\Re(\hat{\qy}_{\rm {DD}})  \ \Im(\hat{\qy}_{\rm {DD}})
\end{bmatrix}^T$,
$\qn=
\begin{bmatrix}
	\Re(\hat{\qw}_{\rm {DD}})  \ \Im(\hat{\qw}_{\rm {DD}})
\end{bmatrix}^T$,
$\qH=
\begin{bmatrix}
	\Re(\hat{\qH}_{\rm {eff}})  & -\Im(\hat{\qH}_{\rm {eff}}) \\ \Im(\hat{\qH}_{\rm {eff}}) & \Re(\hat{\qH}_{\rm {eff}})
\end{bmatrix}$, $\Re (\cdot)$ and $\Im (\cdot)$ take the real and imaginary parts, respectively. $\qy, \qn \in \mathbb{R}^{V\times 1}, \qx \in \mathbb{R}^{U\times 1}$. 
$\qH \in \mathbb{R}^{V\times U}$. We consider the model \eqref{real-y=hx+w} for the rest of the paper.
   
\subsection{GDUNN}
The GDUNN architecture is shown in Fig. \ref{fig_NN}. It consists of $D$ FC layers, and the Tanh activation function is used after each FC layer. 
We extract the interference among DD domain symbols from the estimated DD domain channel matrix $\qH$ in \eqref{real-y=hx+w} and embed it into the last GDUNN layer. Specifically, we use a directed graph to represent the interference, as shown in Fig. \ref{fig_graph}, where nodes feature $x_1,\cdots,x_U$ represent the entries of the transmitted symbol $\qx$ in \eqref{real-y=hx+w}. $e_{i,j}$ is the edge representing the interference between $x_i$ and $x_j$. $e_{i,j} \neq e_{j,i}$ as the interference from $x_i$ to $x_j$ and from $x_j$ to $x_i$ are different. The edge value is defined as $e_{i,j} = \frac{\qh_i^T \qh_j}{\| \qh_j\|^2}$, where $\qh_j$ represents the $j$-th column of channel matrix $\qH$. 
Thus, the adjacency matrix of the graph is expressed as $\qF \in \mathbb{R}^{U \times U}$, where the $i,j$-th entry of $\qF$ is $e_{i,j}$.
The output of GDUNN at iteration $i$ is expressed as, 
\begin{equation}\label{GDUNN_output}
  \hat{\qx}_{\rm GDUNN}^{(i)} = \alpha f_{D}^{(i)}(f_{D-1}^{(i)}(\cdots (f_1^{(i)}(\qz_1^{(i)})))),
\end{equation}
\begin{equation}\label{forward}
    f_d^{(i)}(\qz_d^{(i)}) =   
    \begin{cases} 
    {\rm {Tanh}}(\qW_{d}^{(i)} \qz_d^{(i)} + \qb_{d}^{(i)}),    
    d=1,\dots,D-1,\\
    {\rm {Tanh}}(\qW_{d}^{(i)}(\qF \qz_d^{(i)}) + \qb_{d}^{(i)}),  d=D,
    \end{cases}  
\end{equation}
where $f_{d}^{(i)}(\qz_d^{(i)})\in \mathbb{R}^{O_d \times1}$ and $\qz_d^{(i)}\in \mathbb{R}^{I_d \times1}$ are the output and input of $d$-th FC layer at iteration $i$, respectively. $O_d$ and $I_d$ are the input and output sizes of $d$-th FC layer, respectively. $\qz_1^{(i)}$ is the first FC layer input drawn from a normal distribution $\mathcal{N}(\bold{0}, \bold{1})$ and it is fixed during the iterative process.
$\alpha$ is a constant that restricts the GDUNN output range within the minimum and maximum power of $\tilde{M}$-QAM symbol $\qx$ in \eqref{real-y=hx+w}.  
$\qW_{d}^{(i)} \in \mathbb{R}^{O_d \times I_d}$ and $\qb_{d}^{(i)} \in \mathbb{R}^{O_d \times 1}$ represent the weight matrix and bias vector in $d$-th layer at iteration $i$. Each entry of $\qW_{d}^{(0)}$ and $\qb_{d}^{(0)}$ are initialized following a uniform distribution with a range of $(\frac{-1}{\sqrt{O_d}},\frac{1}{\sqrt{O_d}})$ \cite{7410480}.
 
The stopping criteria in \cite{10225901} is used to control the iterative process of GDUNN by using the computed variance of its output in \eqref{GDUNN_output} for a fixed window size $S$. This is given as 
\begin{align}\label{stop}
	\varsigma^{(i)} = \frac{1}{S} \sum_{j=i-S}^i \| \hat{\qx}_{\rm GDUNN}^{(j)} - \frac{1}{S} \sum_{j'=i-S}^i  \hat{\qx}_{\rm GDUNN}^{(j')} \|^2, i\geq S,
\end{align}
where $\varsigma^{(i)}$ is the variance value at iteration $i$. When $i<S$, the variance calculation is inactive. $S$ determines how many outputs are used for calculating the variance. We compare $\varsigma^{(i)}$ with a threshold $\Psi$. If $\varsigma^{(i)} < \Psi$, the iterative process of GDUNN will stop, and $\hat{\qx}_{\rm GDUNN}^{(i)}, i = T_{\rm GDUNN}$ is the final estimated symbol. Otherwise, we calculate the MSE between the output in \eqref{GDUNN_output} and the received signal $\qy$
\begin{equation}\label{loss}
    \mathcal{L}^{(i)}  = \frac{1}{U} \| \qH  \hat{\qx}_{\rm GDUNN}^{(i)} -  \qy \|^2.
\end{equation}
Adam optimizer \cite{kingma2014adam} is then used to calculate the gradient and update weights $\qW_d^{(i)}$ and biases $\qb_d^{(i)}$ used for iteration $i+1$ by minimizing the calculated error in \eqref{loss}.

\subsection{GDUNN-BPIC} 
In this section, we propose to combine the GDUNN and BPIC OTFS symbol detection.
A brief description of the BPIC algorithm proposed in \cite{9569353} is shown, consisting of Bayesian symbol observation (BSO), Bayesian symbol estimation (BSE), and decision statistics combining (DSC). 
BSO is a matched filter based PIC scheme that is used to estimate the transmitted symbols in iteration $t$ based on the estimated symbols $\hat{\qx}^{(t-1)}$ from the previous iteration, expressed as
\begin{equation}\label{BSO_x}
\mu_q^{(t)} = \hat{x}_q^{(t-1)} + \frac{\qh_q^{T} \left( \qy-\qH \hat{\qx}^{(t-1)} \right)}{\| \qh_q \|^2},
\end{equation} 
where $\mu_q^{(t)}$ is the soft estimate of $q$-th symbol $x_q$ in iteration $t$, $\qh_q$ is the $q$-th column of channel matrix $\qH$. $\hat{\qx}^{(t-1)} =[\hat{x}_1^{(t-1)},\cdots,\hat{x}_q^{(t-1)},\cdots,\hat{x}_{U}^{(t-1)}]^T$ is the vector of the estimated symbol.
The MMSE scheme is used in \cite{9569353} to obtain the initial symbol estimates, as $\hat{\qx}^{(0)}=(\qH^T\qH+\sigma^2\qI)^{-1}\qH^T\qy$ in the first BPIC iteration. Here, we use the output of GDUNN as the initial symbol estimates of the BPIC, i.e., $\hat{\qx}^{(0)}=\Tilde{\qx}_{\rm GDUNN}^{(T_{\rm GDUNN})}$, resulting in the GDUNN-BPIC OTFS detector.
The variance $ \Sigma_q^{(t)} $ of the $q$-th symbol estimate is expressed as
\begin{align}\label{BSO_v}
\Sigma_q^{(t)} &   = \frac{1}{(\qh_q^T \qh_q)^2} \left(\sum_{j=1\atop{j\neq{q}}}^{U}(\qh_q^T \qh_q)^2 v_j^{(t-1)}+(\qh_q^T \qh_q) \sigma^2 \right),
\end{align}
where $v_j^{(t-1)}$ is the $j$-th element in a vector of symbol estimates variance $\qv^{(t-1)}$ in iteration $t-1$ and $\qv^{(t-1)} = [v_1^{(t-1)},\cdots,v_q^{(t-1)},\cdots,v_{U}^{(t-1)}]^T$. We set $\qv^{(0)} = 0$, as the prior knowledge of the variance is unknown at the beginning.
Then the estimated symbol $\qmu^{(t)}=[\mu_1^{(t)},\cdots,\mu_q^{(t)},\cdots,\mu_{U}^{(t)}]^T$and variance $\qSigma^{(t)}=[\Sigma_1^{(t)},\cdots,\Sigma_q^{(t)},\cdots,\Sigma_{U}^{(t)}]^T$ are forwarded to the BSE module.

  
The BSE module computes the Bayesian symbol estimates and the variance of the $q$-th symbol obtained from the BSO module, given as \cite{9569353}
\begin{equation}\label{Bayesian_x}
\hat{x}_q^{(t)} =\mathbb{E} \left[x_q \Big| \mu_q^{(t)} ,\Sigma_q^{(t)} \right] 
\end{equation}
\begin{equation}\label{Bayesian_v}
v_q^{(t)}= \mathbb{E}  \left[ \left| x_q  - \mathbb{E} \left[x_q \Big| \mu_q^{(t)} ,\Sigma_q^{(t)} \right] \right|^{2} \right]. 
\end{equation}
$\hat{x}_q^{(t)} $and $v_q^{(t)}$ are then sent to the following DSC module.

The DSC module performs a linear combination of the symbol estimates and its variance obtained in the BSE model in two consecutive iterations, shown as
\begin{equation}\label{DSC}
\hat{x}_q^{(t)} = \left( 1-\rho_q^{(t)} \right)  \hat{x}_q^{(t-1)}  +   \rho_q^{(t)}   \hat{x}_q^{(t)}
\end{equation}
\begin{equation}\label{DSC_Var}
 v_q^{(t)} = \left( 1-\rho_q^{(t)} \right)  v_q^{(t-1)}  +   \rho_q^{(t)}   v_q^{(t)},
\end{equation}
where $\rho_q^{(t)}= \frac{\epsilon_q^{(t-1)}}{\epsilon_q^{(t)}+\epsilon_q^{(t-1)}}$ is the weighting coefficient, and $\epsilon_q^{(t)}  =  \left\|\frac{\qh_q^{T}}{\| \qh_q\|^2} \left(  \qy - \qH \hat{\qx}^{(t)} \right)\right\|^2.$  
The weighted symbol estimates $\hat{\qx}^{(t)}$  and the variance $\qv^{(t)}$ are then returned to the BSO module for the next iteration. After $T_{\rm BPIC}$ iterations, $\hat{\qx}^{(T_{\rm BPIC})}$ is taken as a vector of symbol estimates. 
\begin{table} 
	\begin{tabular}{| c| c| c|}
		\hline
		
		Detector  & \makecell	{Complexity order} \\
		\hline
	    MMSE\cite{9698103}  &	 $\mathcal{O} (VU^2)$ 	\\
		\hline
        UAMP\cite{9492800} 		 & 	$\mathcal{O}(VU^2+VUT_{\rm UAMP})$ 	 \\
        \hline
        MMSE-BPIC\cite{9569353}  & 	$\mathcal{O}(VU^2+VUT_{\rm BPIC}) $ 	\\
        \hline
        BPICNet\cite{9900413} 	& {$\mathcal{O}(VU^2+VUT_{\rm BPICNet})$ }	 	\\
        \hline
		EP\cite{long2021low} 	 & 	$\mathcal{O}(VU^2T_{\rm EP})$ 	 	\\
        \hline
        DUNN-BPIC \cite{10225901}              & 	 $\mathcal{O} (VUT_{\rm DUNN}+VUT_{\rm BPIC})$ 	\\
		\hline
        GDUNN-BPIC 	                 & 	 $\mathcal{O} (VUT_{\rm GDUNN}+VUT_{\rm BPIC})$ 	\\
		\hline

	\end{tabular}  
	\caption{Computational complexity comparison}
	\label{Tab:complexity}
\end{table}
\section{Computational Complexity Analysis}\label{Complexity}
The computational complexity of GDUNN depends on the FC layers and MSE calculation, which are matrix-vector multiplications. The number of multiplications used for GDUNN is $\mathcal{O} (VUT_{\rm GDUNN})$. The number of multiplications of GDUNN-BPIC is $\mathcal{O} (VUT_{\rm GDUNN}+VUT_{\rm BPIC})$.
Table \ref{Tab:complexity} shows the computational complexity order in terms of the number of multiplications for different OTFS detectors, where $T_{\rm UAMP}, T_{\rm BPIC}$, and $T_{\rm EP}$ represent the iterations needed for the UAMP, MMSE-BPIC, and EP, respectively. 
Our proposed GDUNN-BIPC has lower computational complexity orders than MMSE, MMSE-BPIC, UAMP, and EP. 
Here, we give a numerical example for $M=N=8,l_{max}=2,k_{max}=1, T_{\rm GDUNN}=100$, and $T_{\rm BPIC}=T_{\rm UAMP}=T_{\rm EP}=10$ as suggested in \cite{9569353}, \cite{9900413}, \cite{9492800}, and \cite{long2021low}, respectively. Based on Table \ref{Tab:complexity}, GDUNN-BPIC is less complex than \cite{9569353,9900413,9492800}, and it is approximately 10 times less complex than \cite{long2021low}.

\begin{figure}
\centering
\subfigure[SER performance regarding window size $S$ ]
{\includegraphics[width=0.9\textwidth]{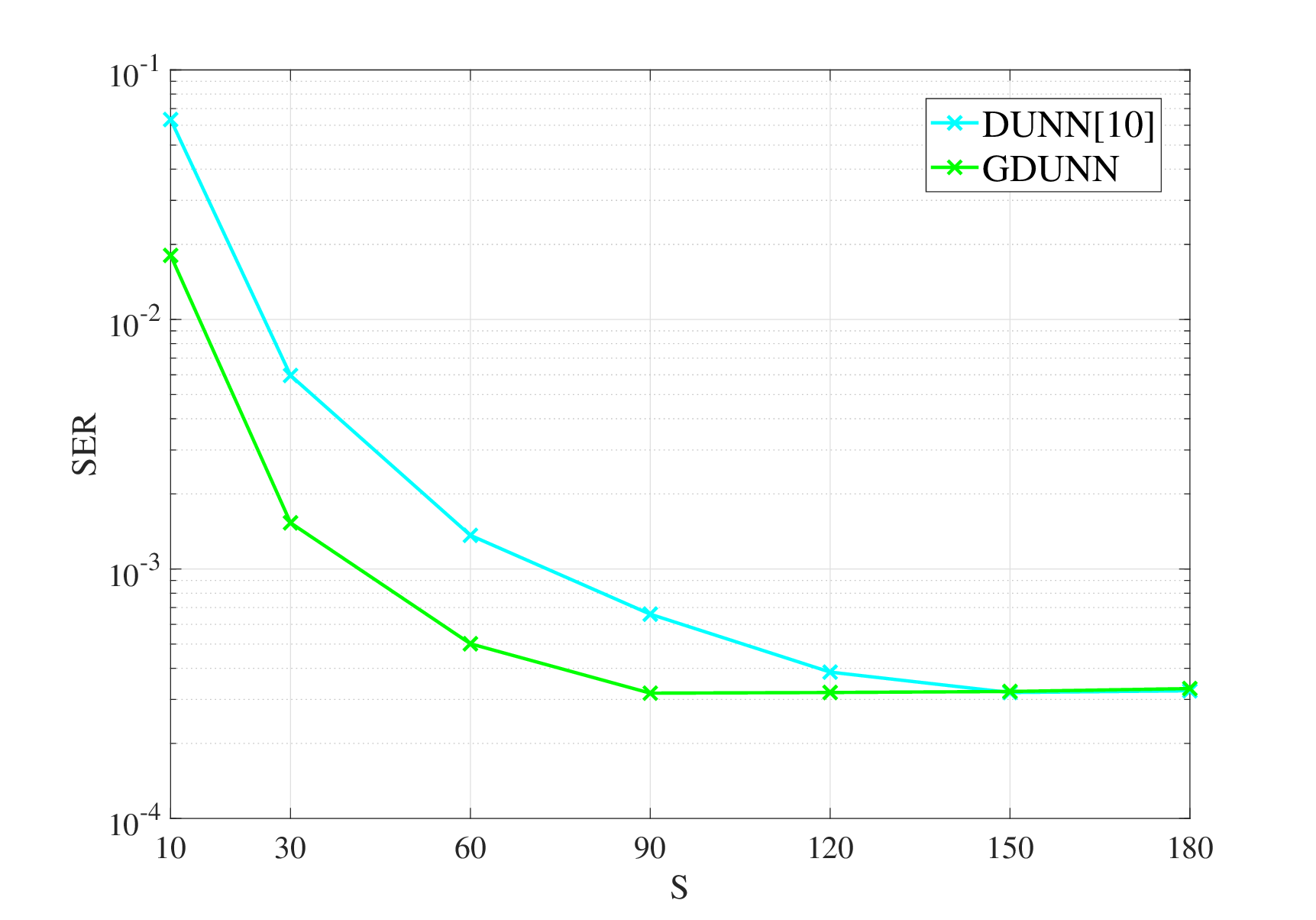}\hfill
\label{fig_s}}
\subfigure[CDF of iteration numbers]
{\includegraphics[width=0.9\textwidth]{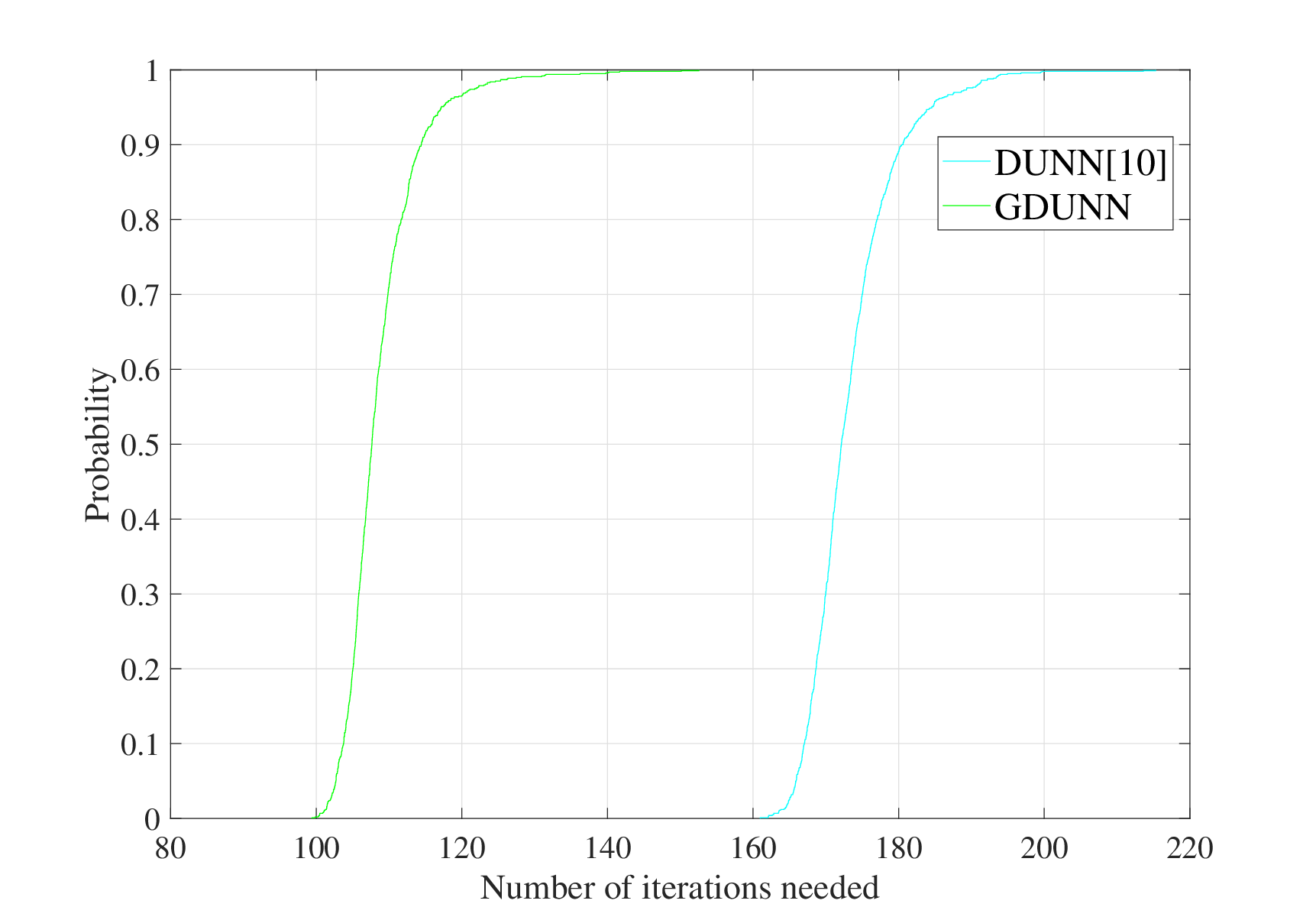}\hfill
\label{fig_cdf}}
\caption{Convergence comparison of DUNN and GDUNN for $SNR=28$ dB }
\end{figure}


\begin{figure}
\centering
\subfigure[SER vs SNR]
{\includegraphics[width=0.9\textwidth]{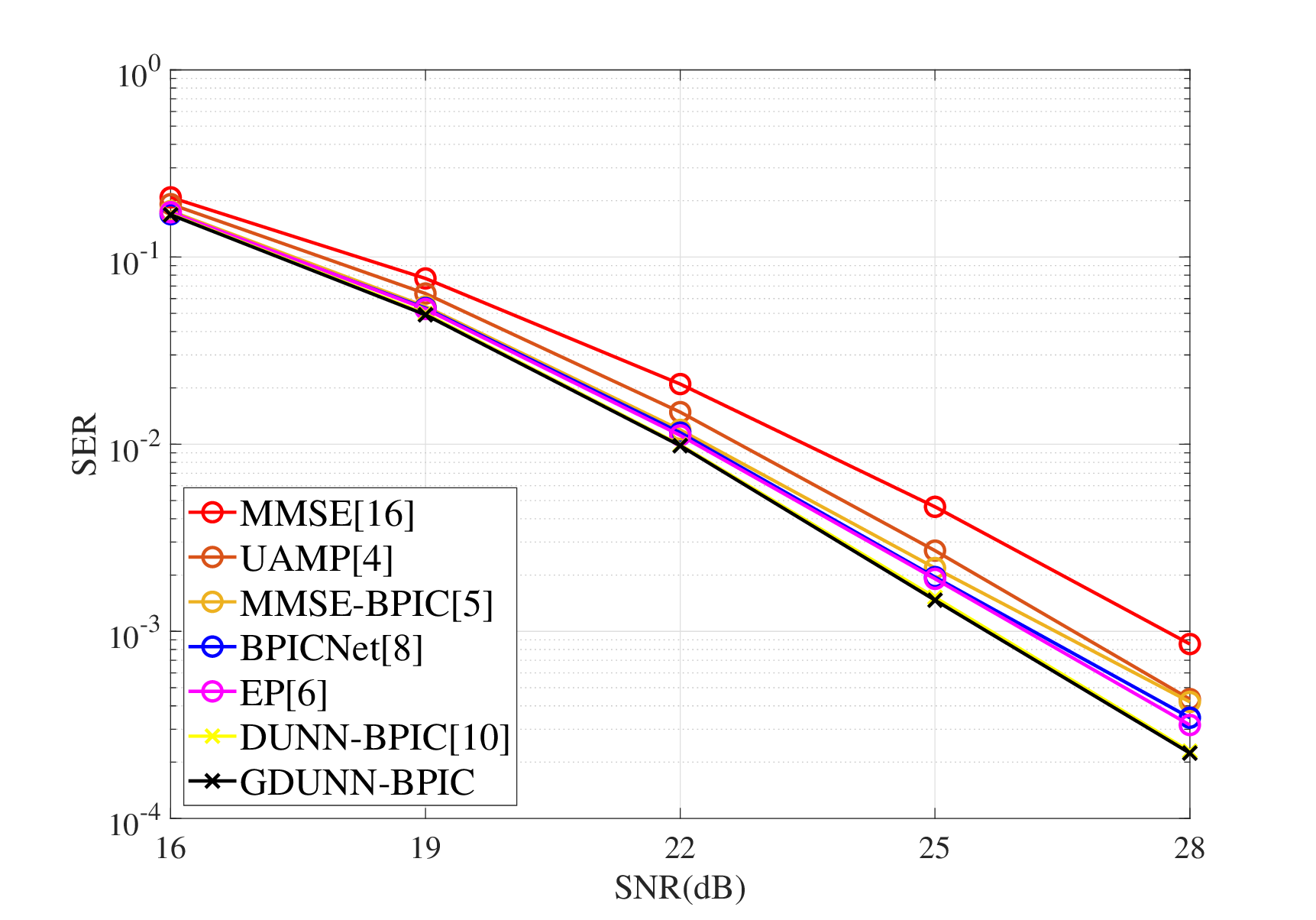}\hfill
\label{fig_ser}}
\subfigure[SER vs pilot power $P_p$]
{\includegraphics[width=0.9\textwidth]{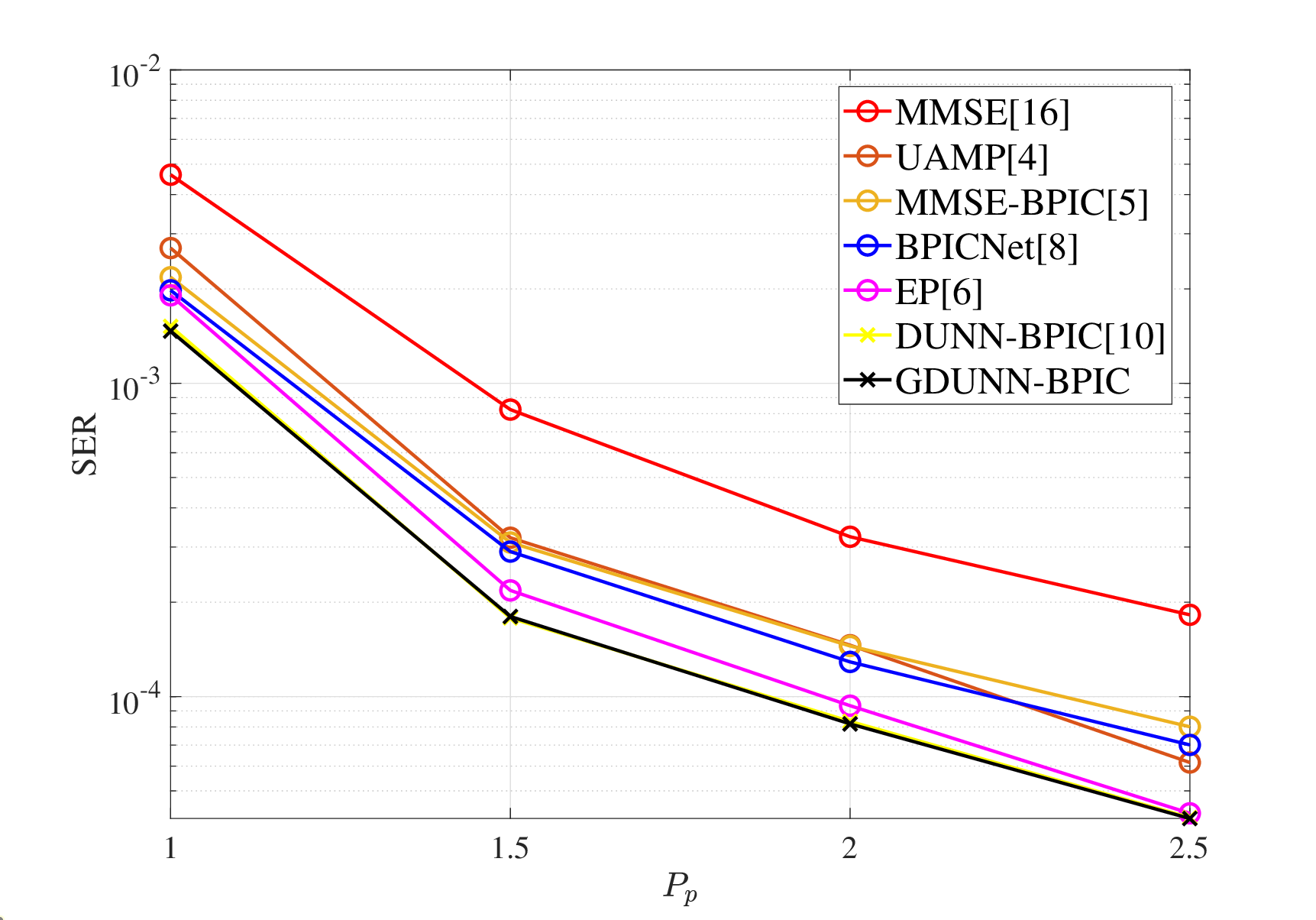}\hfill
\label{fig_ser_p}}
\caption{SER performance comparison of OTFS detectors}
\end{figure}
  


\section{Numerical results}\label{Results}
In this section, we first analyze the convergence of the proposed GDUNN in comparison with DUNN \cite{10225901}. We then compare the SER performance of our proposed GDUNN-BPIC with state-of-the-art ones, including MMSE \cite{9698103}, UAMP\cite{9492800}, MMSE-BPIC \cite{9569353}, BPICNet \cite{9900413}, EP\cite{long2021low}, and DUNN-BPIC \cite{10225901} under imperfect CSI. 
In the simulations, we consider $\Tilde{M}=16$, $N=M=8, l_{max}=2, k_{max}=1, P=4$. Subcarrier spacing and carrier frequency are 15 KHz and 6 GHz, respectively, resulting in a maximum tolerable mobility of 337 Km/h. The pilot is put in the position of $(k_p,l_p)=(3,3)$ and $P_p=1$.  
Note that the transmitted symbols have a unit average power, and therefore, we set $\alpha=\frac{3}{\sqrt{10}}$.
The learning rate of GDUNN is set to 0.01, and $I_d=4,8,16,32,U$ and $O_d=8,16,32,U,U$ for $d=1,\dots D$, respectively, and $D=5$. By deleting the last GDUNN layer (i.e., $d=D$ in \eqref{forward}), we can get DUNN as in \cite{10225901}.
$T_{\rm BPIC}=T_{\rm UAMP}=T_{\rm BPICNet}=T_{\rm EP}=10$. The imperfect CSI is considered for all the simulations below.

Fig. \ref{fig_s} shows the SER performance of DUNN and GDUNN regarding window size $S$ with threshold $\Psi=0.001$ in \eqref{stop}. Fig. \ref{fig_s} demonstrates that GDUNN and DUNN achieve the best SER performance with minimum window sizes $S=90$ and $S=150$, respectively.  
Fig. \ref{fig_cdf} shows the cumulative density function (CDF) of the number of iterations needed for DUNN and GDUNN using the minimum window sizes. It demonstrates that GDUNN needs approximately $40\%$ fewer iterations than DUNN. Therefore, the proposed GDUNN-BPIC is less complex than DUNN-BPIC.
Fig. \ref{fig_ser} shows the SER performance of the proposed OTFS detector, where GDUNN-BPIC achieves a similar SER performance to DUNN-BPIC and outperforms the others. 
In Fig. \ref{fig_ser_p}, we fix the $SNR=25$ dB and increase the power of the pilot $P_p$, resulting in a decrease in CSI estimate error, as illustrated in \cite{8671740}. 
Fig. \ref{fig_ser_p} demonstrates that the proposed GDUNN-BPIC still achieves the best SER performance among all the state-of-the-art OTFS detectors when the error of CSI estimates decreases. Therefore, the proposed GDUNN-BPIC is robust to imperfect CSI estimates compared to others.

\section{Conclusion}
In conclusion, we proposed a graph-based decoder UNN OTFS detector in the presence of imperfect CSI, referred to as GDUNN-BPIC. Simulation results showed that GDUNN needs fewer iterations than DUNN due to the embedding of prior knowledge. The proposed GDUNN-BPIC outperforms state-of-the-art OTFS detectors using lower computational complexity under imperfect CSI.


{\renewcommand{\baselinestretch}{1.1}
	\begin{footnotesize}
		\bibliographystyle{IEEEtran}
		\bibliography{IEEEabrv,paper}
\end{footnotesize}}

\end{document}

%% file: macro.TeX
    \def\Complex{{\rm\rule[.23ex]{.03em}{1.1ex}\kern-.3em{C}}}

    \newcommand{\be}{\begin{equation}} \newcommand{\ee}{\end{equation}}
    \newcommand{\bea}{\begin{eqnarray}} \newcommand{\eea}{\end{eqnarray}}
    \newcommand{\benum}{\begin{enumerate}} \newcommand{\eenum}{\end{enumerate}}

    \newcommand{\qa}{{\bf a}}
        \newcommand{\qb}{{\bf b}}

        \newcommand{\qh}{{\bf h}}

        \newcommand{\qn}{{\bf n}}

        \newcommand{\qr}{{\bf r}}
        \newcommand{\qs}{{\bf s}}

        \newcommand{\qv}{{\bf v}}
        \newcommand{\qw}{{\bf w}}
        \newcommand{\qx}{{\bf x}}
        \newcommand{\qy}{{\bf y}}
        \newcommand{\qz}{{\bf z}}
    
        \newcommand{\qA}{{\bf A}}

        \newcommand{\qF}{{\bf F}}
        \newcommand{\qG}{{\bf G}}
        \newcommand{\qH}{{\bf H}}
        \newcommand{\qI}{{\bf I}}

        \newcommand{\qR}{{\bf R}}

        \newcommand{\qW}{{\bf W}}
        \newcommand{\qX}{{\bf X}}
        \newcommand{\qY}{{\bf Y}}

        \newcommand{\qDelta}{{\boldsymbol \Delta}}

        \newcommand{\qSigma}{{\boldsymbol \Sigma}}

        \newcommand{\qmu}{{\boldsymbol \mu}}
